# Robust isothermal electric switching of interface magnetization: A route to voltage-controlled spintronics

Xi He[1], Yi Wang[1], Ning Wu[1], Anthony Caruso[2], Elio Vescovo[3], Kirill D. Belashchenko[1], Peter A. Dowben[1], Christian Binek[1,*]

[1]*Department of Physics & Astronomy and the Nebraska Center for Materials and Nanoscience, University of Nebraska, Lincoln, NE, 68588-0111, USA*

[2]*Department of Physics, 257 Flarsheim Hall, University of Missouri, 5110 Rockhill Road, Kansas City KS 64110, USA*

[3]*Brookhaven National Laboratory, National Synchrotron Light Source, Upton, NY, 11973, USA*

[*]To whom correspondence should be addressed. E-mail: cbinek2@unl.edu

**Spintronics strives to exploit the spin degree of freedom of electrons for an advanced generation of electronic devices[1,2]. In particular, voltage-controlled spin electronics is of vital importance to continue progress in information technology. The major objective of such an advanced technology is to reduce power consumption while enhancing processing speed, integration density, and functionality in comparison with present day complementary metal-oxide semiconductor (CMOS) electronics[3,4,5,6]. Almost all existing and prototypical solid-state spintronic devices rely on tailored interface magnetism, enabling spin-selective transmission or scattering of electrons. Controlling magnetism at thin-film interfaces, preferably by purely electrical means, is a key challenge to better spintronics[7,8,9,10]. This letter presents compelling evidence of a roughness-**

**insensitive and electrically controllable ferromagnetic state at the (0001) surface of antiferromagnetic chromia. We place such a ferromagnetic surface in close proximity with a ferromagnetic Co/Pd multilayer film. Quantum mechanical exchange coupling at the interface between chromia and Co/Pd induces an electrically controllable unidirectional magnetic anisotropy in the Co/Pd film. This electrically controlled exchange bias effect allows for reversible isothermal shifting of the global magnetic hysteresis loop of the Co/Pd film along the magnetic field axis from negative to positive values. From a technological vantage point, optimized functionality of spintronic applications requires non-volatile control of the magnetization on a mesoscopic scale. Hence, this voltage controlled robust macroscopic magnetization reversal at room temperature can be considered a paradigm shift away from complex oxides revealing the importance of the simple antiferromagnetic chromia for spintronic applications.**

If there is such a thing as the holy grail of spintronics, electric control of surface and interface magnetism is certainly a candidate. The significance of controlled interface magnetism started with the exchange bias effect exploited by most present day spintronic devices. Exchange bias is a coupling phenomenon at magnetic interfaces which reflects its presence most prominently by a shift of the ferromagnetic hysteresis along the magnetic field axis and is quantified by the amount $\mu_0 H_{EB}$ of the shift[11]. The exchange bias pinning of ferromagnetic thin films is extensively applied in giant magnetoresistance (GMR) and tunnelling magnetoresistance (TMR) structures of magnetic field sensors and modern magnetic read heads[12]. The latter enabled the ongoing exponential growth of the areal data storage density in magnetic hard disk drives.

Electric control of exchange bias has been proposed for various spintronic applications which go beyond GMR and TMR technology[5]. Voltage control of magnetism is preferable over current-induced switching which, despite tremendous progress, remains inherently power consumptive[1]. One approach to such voltage control hinges on the reversible, laterally uniform, isothermal electric tuning of the exchange bias field at room temperature, which remains a major challenge.

Early attempts in electrically controlled exchange bias tried to exploit the linear magnetoelectric susceptibility of the antiferromagnetic material $Cr_2O_3$ as an active exchange bias pinning system[13]. In a magnetoelectric material an applied electric field induces a net magnetic moment which can be used to electrically manipulate the magnetic states of an adjacent exchange coupled ferromagnetic film[14]. The small value of the maximum parallel magnetoelectric susceptibility $\alpha_{me}^{\|}(T=263K) \approx 4.13$ ps/m of $Cr_2O_3$[15] led many researchers to the conclusion that multiferroic materials are better suited for this purpose. Such materials have two or more ferroic order parameters, such as ferroelectric polarization and (anti)ferromagnetic order[16].

The potential for an increased magnetoelectric response, for the multiferroic materials, was dictated by the maximum possible value of $\alpha_{me}^{ij}$. It is determined by the geometric mean of the ferroic susceptibilities, both of which can individually be very high in multiferroics[17, 18, 19, 20]. Coupling between these order parameters has been demonstrated[21]. However, it is typically weak, and the theoretical upper limit of $\alpha_{me}^{ij}$ is rarely reached[16]. The most promising multiferroic single-phase materials used for electrically controlled exchange bias are $YMnO_3$ and $BiFeO_3$[22, 23]. Complete suppression of the exchange bias has been achieved at $T$ = 2K in an $YMnO_3$/NiFe (permalloy) heterostructure. However, this effect is irreversible. Moreover, the limitation to low temperatures makes YMnO$_3$ unsuitable for applications. The situation is better with BiFeO$_3$. In BiFeO$_3$/CoFe heterostructures, local magnetization reversal on a lateral length scale of up to 2 µm has been demonstrated[23, 24]. However, global magnetization reversal, which could be revealed in macroscopic magnetic hysteresis, has not been achieved, because only the domain walls in BiFeO$_3$ are responsible for the local control of pinning effects. Global, but not isothermal magnetoelectric switching has been achieved in the pioneering Cr$_2$O$_3$/CoPt heterostructure [13]. However, each sign reversal of the exchange bias field required a new magnetoelectric annealing procedure, in which the pinning layer is cooled from $T > T_N$ to $T < T_N$ in the presence of both electric and magnetic fields. Isothermal electric control of exchange bias has been attempted by various groups, but with only marginal success[25, 26]. The result is that reversible and global electrically controlled exchange bias performed isothermally at room temperature remained elusive.

This letter reveals an unconventional ferromagnetism at the (0001) surface of the magnetoelectric antiferromagnet Cr$_2$O$_3$ and the stunning consequences for electrically controlled exchange bias and magnetization. New insights were achieved by combining first-principles calculations of the spin-resolved density of states, measurements of the spin-resolved photoemission, and magnetometry at Cr$_2$O$_3$ (0001) surfaces and interfaces in exchange bias heterostructures. Based on the understanding of the surface ferromagnetism of Cr$_2$O$_3$ (0001), a new concept of Cr$_2$O$_3$ (0001)-based exchange bias is implemented. As a result, a reversible, isothermal, and global electric control of exchange bias is demonstrated at room temperature, evidenced by reproducible electrically-induced discrete shifts of the global magnetic hysteresis loops along the magnetic field axis.

Magnetically uncompensated surfaces of antiferromagnetically ordered single crystals have been subject of intense investigations in particular in the framework of exchange bias[11]. The surface magnetization of an uncompensated antiferromagnetic surface with roughness averages out, so that only a small statistical fluctuation remains for exchange coupling with the adjacent ferromagnet[27].

The (0001) surface of $Cr_2O_3$ is a remarkable exception. As illustrated in Fig. 1a, a rare combination of antiferromagnetic ordering and surface termination creates an unusual situation. Each antiferromagnetic domain has an uncompensated surface magnetic monolayer with aligned moments on all surface $Cr^{3+}$ ions, even if the surface is not atomically flat. Two features conspire to produce this property. First, the corundum lattice of $Cr_2O_3$ can be imagined as a layered arrangement of buckled $Cr^{3+}$ ions sandwiched between the triangular layers of $O^{2-}$ ions (best seen in Fig. 1b and 1c)[28]. The electrostatically stable charge-neutral surface of this crystal is terminated by a semi-layer of Cr; this termination can be viewed as the cleavage of the crystal in the middle of the buckled $Cr^{3+}$ layer[29]. Second, Cr ions which are structurally similar with respect to the underlying O layer have parallel spins in the antiferromagnetically ordered state. As a result, a single-domain antiferromagnetic state has all surface Cr spins pointing in the same direction. Note that while we have shown the surface Cr ions in bulk-like positions in Fig. 1, this assumption is immaterial for the existence of the surface magnetization, as long as Cr preferentially occupies a particular site at the surface.

In single-crystalline $Cr_2O_3$ the antiferromagnetic order appears in the form of two 180º antiferromagnetic domains (see Fig. 1 and Supplementary Fig.1)[14]. These two domains have surface magnetic moments of opposite sign. If the degeneracy of the two domain types is not lifted, the system develops a random multi-domain state with zero net surface magnetization when it is cooled below $T_N$. However, magnetoelectric annealing selects one of these 180° domains by exploiting the free energy gain $\Delta F = \alpha\, EH$ [14]. As a result, even a rough $Cr_2O_3$ (0001) surface becomes spin-polarized when an antiferromagnetic single-domain state is established. Evidence of this roughness-insensitive surface magnetism is revealed by magnetometry (Supplementary Fig. 2 and discussion) as well as spin and energy-resolved X-ray photoemission spectroscopy (XPS). Interpretation of the latter is supported by calculations of the site-resolved density of states revealing a spin-polarized surface band above the valence band maximum in agreement with experimental findings. The XPS performed on one of our MBE-grown $Cr_2O_3$ (0001) samples is sensitive to occupied surface electronic states.

Fig. 2a shows the photoelectron intensity versus binding energy measured at 100 K. Here the $Cr_2O_3$ (0001) thin film has been cooled from $T > T_N$ in a small magnetic field of 30mT alone, into a multi-domain antiferromagnetic state. Spin-up and spin-down photoelectron intensities $I_{\uparrow,\downarrow}$ (red circles and blue squares) are virtually identical, indicating negligible net surface magnetization and polarization. However, the signal is clearly spin-split after magnetoelectric annealing in $E = 3.85 \times 10^{-4}$ kV/mm and $\mu_0 H = 30$ mT (compare red circles and blue squares in Fig. 2b), demonstrating high net spin polarization at the surface. The spin contrast $P = (I_\uparrow - I_\downarrow)/(I_\uparrow + I_\downarrow)$, exhibited by triangles in Fig. 2b, is seen to increase significantly close to the valence band maximum, $E_F$, (green triangles). We identify this feature with the contribution from the spin-polarized surface states of Cr 3d character. To corroborate this interpretation we decompose the spin dependent photoemission spectra $I_{\uparrow,\downarrow}$ into contributions from Cr 3d bulk and surface states. The contribution above $E_F$ (green) is interpreted as an additional spin-polarized surface state.

This interpretation is in accordance with our first principles calculation of the layer resolved density of states (DOS) displayed in the inset of Fig. 2a. The DOS of a representative central layer with spin-up sublattice (majority/minority in red/blue) magnetization is displayed by the lower two curves of the inset. The DOS of the surface layer is displayed by the two upper curves in the inset of Fig. 2a. Note that in addition to the bulk states a surplus spin-up density of states (green) appears above the valence band maximum. This is consistent with our experimental findings in photoemission (Fig. 2b) and magnetometry (Supplementary Fig. 2 and discussion).

Experimental and theoretical evidence together point very strongly to the existence of a roughness-insensitive ferromagnetic state at the $Cr_2O_3$ (0001) surface when the underlying $Cr_2O_3$ single crystal is in an antiferromagnetic single domain state. This ferromagnetic surface moment can be isothermally switched by electrical means giving rise to reversible switching of large exchange bias fields in our perpendicular exchange bias heterostructure $Cr_2O_3(0001)/Pd0.5nm/(Co0.6nmPd1.0nm)_3$.

Fig. 3 demonstrates large isothermal electric switching of the exchange bias field. It is achieved by leaving the realm of the linear magnetoelectric effect. The latter gives rise to a miniscule electric control effect only[25, 26]. In contrast to this small linear effect, significant electrically controlled switching requires to overcome a critical threshold given by the product $|EH|_c$ where $E$ and $H$ are isothermally applied axial electric and magnetic fields. Initially the heterostructure has been magnetoelectrically annealed in $EH$ > 0 with $E$ = 0.1 kV/mm and $\mu_0 H$ = 77.8 mT down to $T$ = 303 K. The hysteresis loops are measured isothermally at $T$ = 303 K and $E$ = 0. Red squares show the virgin loop with positive exchange bias of $\mu_0 H_{EB} = +6$ mT. Next, without changing the temperature, a field product $EH < 0$ of individual fields $E$ = 2.6 kV/mm and $\mu_0 H$ = -154 mT is applied and followed by a measurement of the hysteresis loop in $E$ = 0. Blue circles show the resulting loop with a pronounced negative exchange bias of $\mu_0 H_{EB} = -13$ mT. The isothermal switching of the exchange bias field implies a field-induced switching of the antiferromagnetic single-domain state of $Cr_2O_3$ into the opposite antiferromagnetic registration. This switching is accompanied by a reversal of the interface magnetization. The upper inset of Fig. 3 displays a sequence of switched exchange bias fields obtained by switching the electric field back and forth between $E$ = + 2.6 kV/mm and $E$ = -2.0 kV/mm at constant set field $\mu_0 H$ = -154 mT, all at a constant temperature $T$ = 303 K. The reproducible switching shows no signs of aging. The asymmetry between positive and negative exchange bias values is a consequence of a difference in the interface magnetization $S_{Cr_2O_3}$ for negative and positive exchange bias.

A nonlinear magnetoelectric switching of the antiferromagnetic single domain state of $Cr_2O_3$ was reported as far back as 1966 by Martin and Anderson[30]. Their work discussed, in detail, that the isothermal switching between the two different antiferromagnetic registrations of the $Cr_2O_3$ domains is possible if sufficiently strong field products $E \cdot H$ are applied along the $c$-axis[30]. This switching is a thermally activated process. At constant temperature there is a critical value $|EH|_c$, above which the system settles in the single-domain state with the lowest free energy, even if this requires a switching of the entire antiferromagnetic spin structure. This hysteretic switching of $Cr_2O_3$ is directly reflected in the hysteresis of the electric-field dependence of the exchange bias field.

A different view on these experiments indicates the route towards stable macroscopic magnetization reversal by electric means at room temperature. Here we have switched the remanent magnetization, $M_r$, by electric means, without changing the sign and magnitude of the applied magnetic field. Roughly a 50% change of the remanent magnetization between $E=+2.6$kV/mm and $E=-2.0$kV/mm is demonstrated by the lower inset in Fig. 3. Note that the range between the magnetic set field and zero field can be made arbitrarily small when taking advantage of the finding $|EH|_c = const$ discussed below in detail.

Fig. 4 displays the threshold character of electric switching at $T = 303$ K. All data are taken after magnetoelectric annealing in $E = 0.1$ kV/mm and $\mu_0 H = 77.8$ mT. The hysteretic electric field dependence, $\mu_0 H_{EB}$ vs. $E$, is determined from individual magnetic hysteresis loops measured in $E=0$. Each data point results from a loop measured after isothermal exposure of the sample to one of various $E$-fields and fixed magnetic field $\mu_0 H = -115$ mT (circles), $\mu_0 H = -154$ mT (triangles), and $\mu_0 H = -229$ mT (squares), respectively. Two major characteristics are observed in the $\mu_0 H_{EB}$ vs. $E$ data. First, for a given positive magnetic field there is a critical negative and positive electric field, $E_c$, where switching of the exchange bias field takes place. The rectangular hysteresis $\mu_0 H_{EB}$ vs. $E$ is in perfect agreement with the isothermal switching of the antiferromagnetic domain state of $Cr_2O_3$ reported in (ref. 30). This includes details such as the asymmetry between the negative and the positive switching field.

The insets (a) and (b) of Fig. 4 show that the critical switching fields of the exchange bias obey the relation $|EH|_c = \text{const}$ corresponding to the switching of the $Cr_2O_3$ antiferromagnetic single domain[30]. Solid squares are data points of $E_c$ for magnetic fields $\mu_0 H = $ -115, -154 and -229 mT. The lines are fits of the functional form $H = \text{const}/E_c$. This shows that the switching effect originates from the coherent flip of the antiferromagnetic registration of the $Cr_2O_3$ pinning system. The inversion of the antiferromagnetic spin structure is accompanied by the reversal of the $Cr_2O_3$ (0001) interface magnetization which in turn causes switching of the exchange bias field. Since this switching is induced at a threshold of the product $|EH|_c$, the *H*-field can be made arbitrarily small when *E* is scaled up accordingly resulting in pure electric switching between distinct remanent magnetization states $M_r$ (see lower inset Fig. 3). There is plenty of room for *E*-field increase by shrinking the thickness of the pinning layer down to the nanoscale. The stray magnetic field of the perpendicular anisotropic ferromagnetic top layer may already be sufficient to enable pure voltage control of the exchange bias field, along with numerous potentially revolutionizing spintronic applications[1, 5].

Electric control of magnetism, at room temperature, is the basis of advanced spintronics for post-CMOS technology. The intensive research efforts on multiferroic materials of recent years are to a large extent driven by the possibilities of electrically controlled magnetism at room temperature. We have shown in a combination of experiment and theory that $Cr_2O_3$, the archetypical magnetoelectric antiferromagnet, revives as an outstanding candidate for reversible electric control of magnetism at room temperature. This control is made possible by the peculiar roughness-insensitive ferromagnetic spin termination at the (0001) surface of $Cr_2O_3$. In the highly nonlinear magnetoelectric regime the antiferromagnetic order can be electrically switched along with the surface magnetization. This phenomenon takes place at the interface between $Cr_2O_3$ (0001) and a ferromagnetic Co/Pd multilayer film. In this perpendicular exchange bias system, a reversible and global electric switching of the exchange bias field was realized isothermally at room temperature. This observation opens up exciting prospects for spintronic applications.

**Acknowledgements** This work is supported by NSF through Career DMR-0547887, by the Nebraska Research Initiative (NRI), by the MRSEC Program of the NSF, and by the NRC/NRI supplement to MRSEC. K.D.B. is a Cottrell Scholar of Research Corporation. Technical help from S.-Q. Shi in the calculation of DOS is acknowledged. We are thankful to Crystal GmbH for providing excellent $Cr_2O_3$ single crystals.

**Figures and Legends**

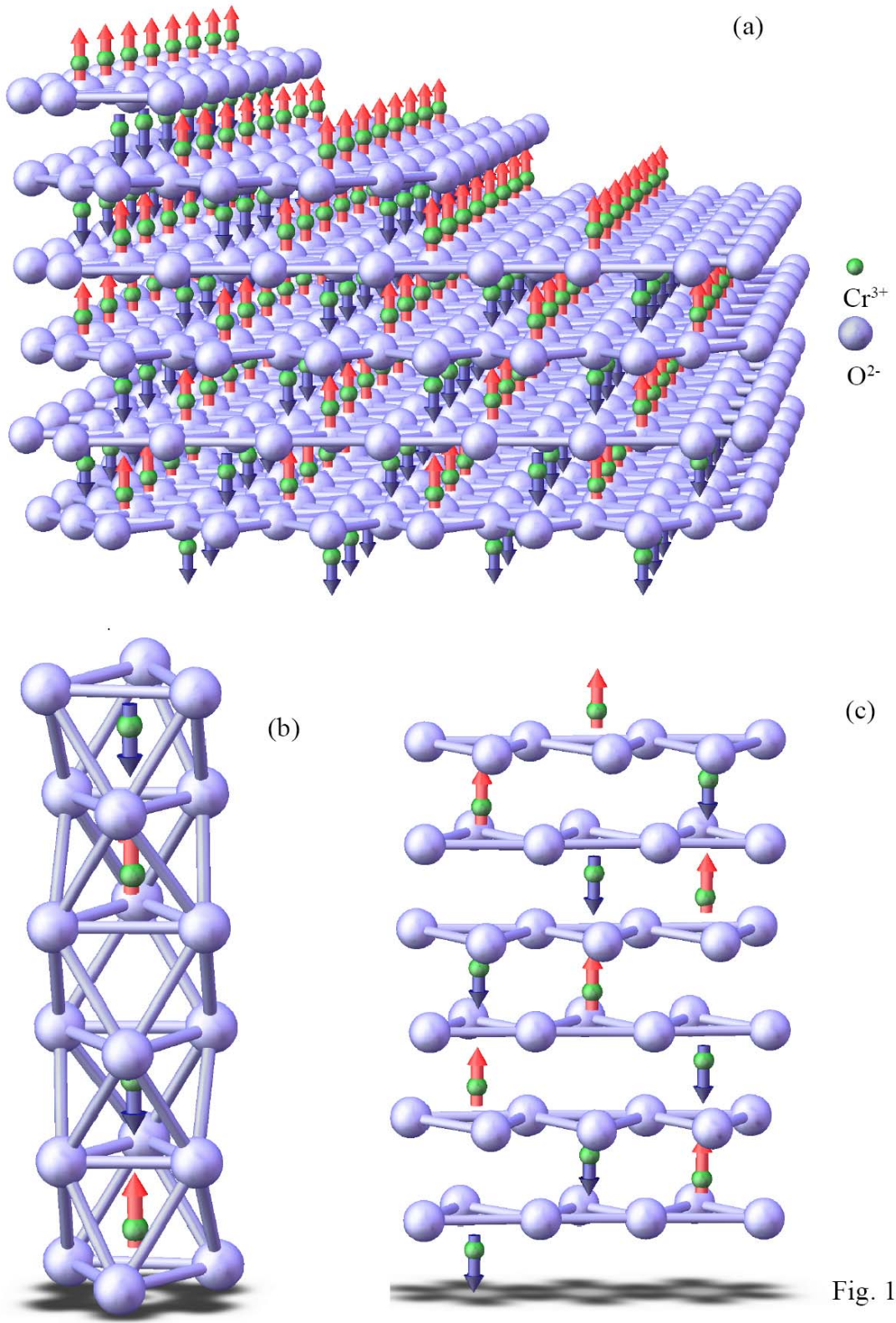

Fig. 1

Fig. 1: (a) The spin structure of a $Cr_2O_3$ single crystal with a stepped (0001) surface is shown for one of its two antiferromagnetic single domain states. Up (red) and down (dark blue) spins of the $Cr^{3+}$ ions (green spheres) point along the *c*-axis. (b) Light blue spheres represent the $O^{2-}$ ions which are arranged in chains of face-sharing distorted octahedra surrounding the $Cr^{3+}$ ions. (c) Shows a more detailed view of the buckled arrangement of the $Cr^{3+}$ ions sandwiched between adjacent oxygen layers.

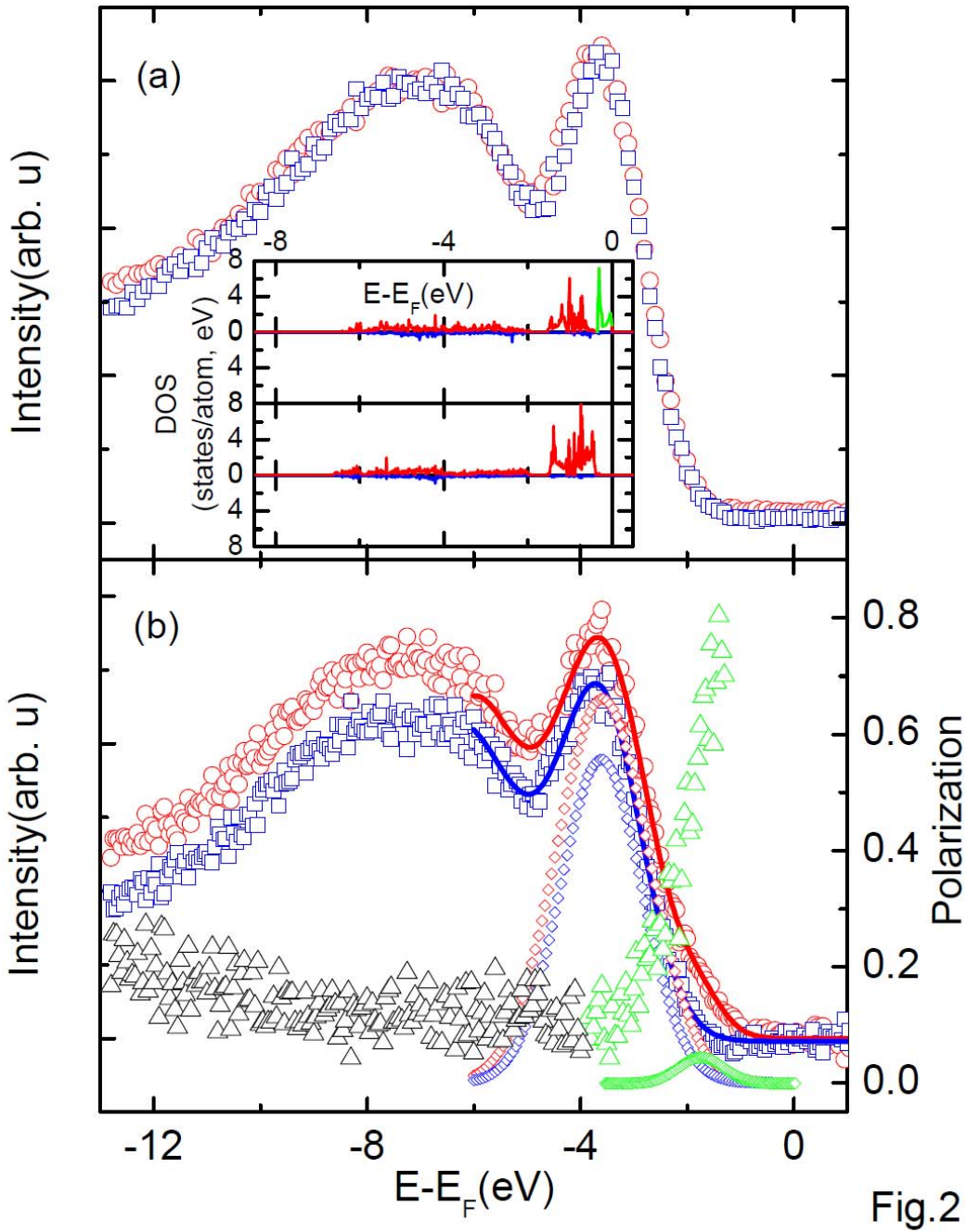

Fig. 2: (a) Shows the intensity of photo-electrons versus binding energy from a $Cr_2O_3$ (0001) surface measured at $T = 100$ K after cooling in $\mu_0 H = 30$ mT and $E=0$ from $T >$

$T_N$. Spin-up and spin-down intensities are displayed by red circles and blues squares, respectively. Inset in (a) shows the result of a first principles calculation of the layer resolved DOS. Color code follows experiment. Green line indicates a surplus surface state with spin-up polarization. (b) Shows spin-up (red circles) and spin-down (blue squares) intensities after magnetoelectric annealing in $E = 3.85\times10^{-4}$ kV/mm and $\mu_0H = 30$ mT. Lines are best fits of multiple-peak Gaussian functions. Diamonds show Cr 3d spin-up (red and green) and spin-down (blue) contributions extracted from the fits. The Gaussian fit displayed by green diamonds reflects specific surface states. Color code matches the theoretical DOS data. Triangles show the contrast, $P$, in the spin dependent intensities versus binding energy. Green triangles highlight the contribution from the surface state.

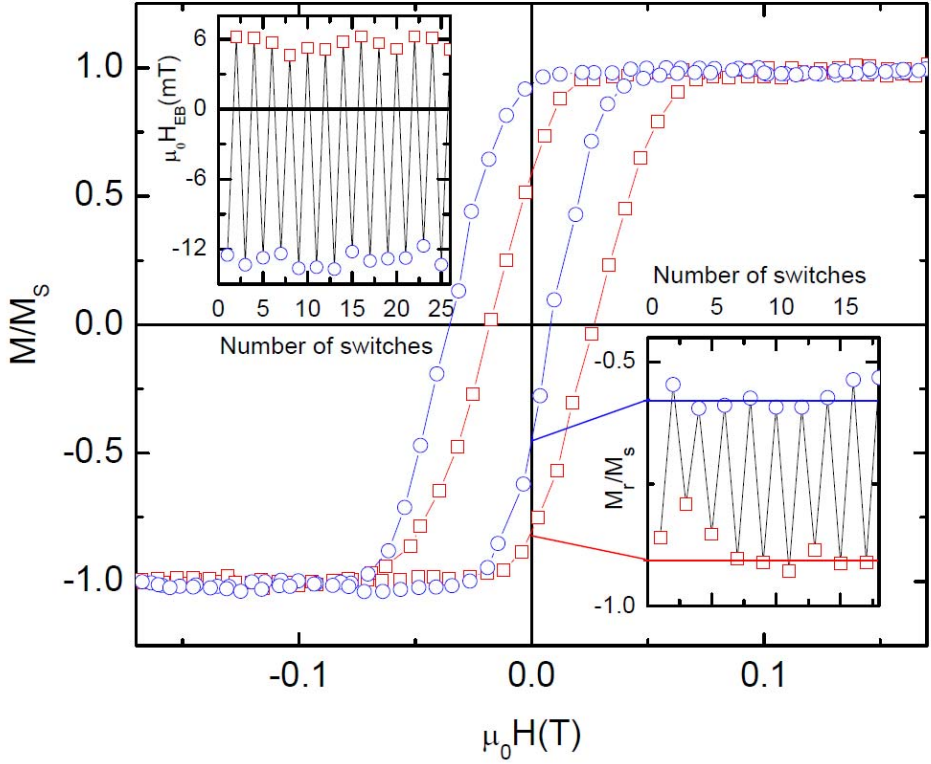

Fig.3

Fig. 3: Isothermal electric switching of the exchange bias field in $Cr_2O_3$ (0001)/ Pd0.5nm/(Co0.6nmPd1.0nm)$_3$ at $T$ = 303 K after initial magnetoelectric annealing in $E$ = 0.1 kV/mm and $\mu_0H$ = 77.8 mT. Hysteresis loops are measured by polar Kerr magnetometry in $E$ = 0, respectively. Red squares show the virgin curve with a positive exchange bias field of $\mu_0H_{EB}$ = +6 mT. Isothermal field exposure in $E$ =+2.6 kV/mm and $\mu_0H$ =-154 mT gives rise to a loop with a negative exchange bias field of $\mu_0H_{EB}$ = -13 mT (blue circles). Upper inset shows $\mu_0H_{EB}$ vs. number of repeated isothermal switching through exposure to $E$ =+ 2.6kV/mm (blue circles) and $E$ =-2 kV/mm (red squares) at constant $\mu_0H$ = -154 mT, respectively. Lower inset shows $M_r/M_s$ vs. number of repeated isothermal switching under the same conditions which apply to the upper inset.

Horizontal lines (red and blue) indicate the average change in $M_r/M_s$ for electrically switched AF states.

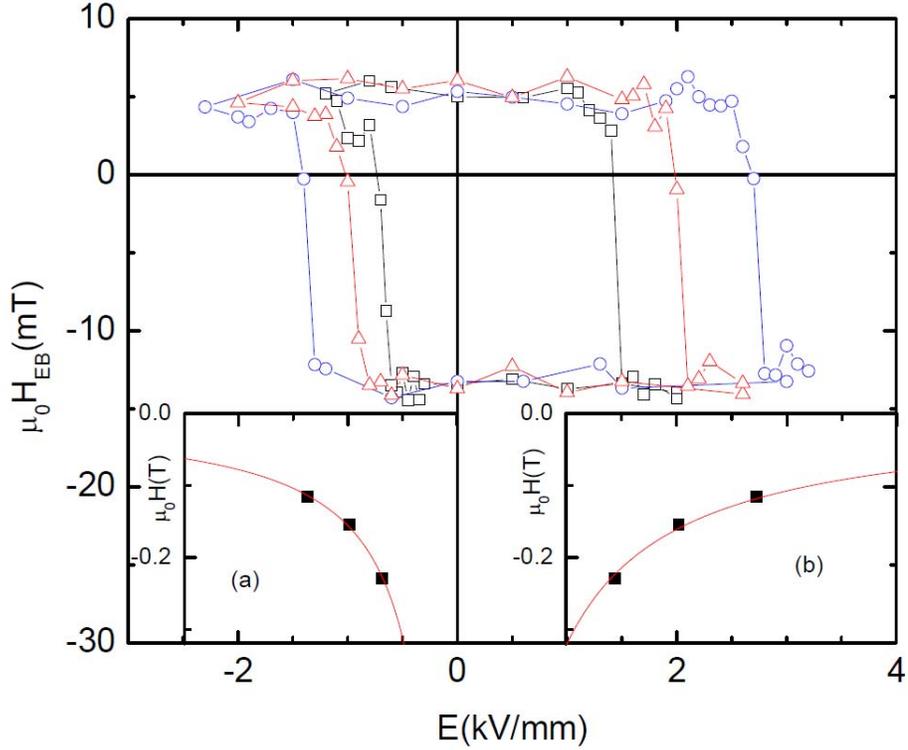

Fig. 4: Hysteretic electric field dependence of the exchange bias field, $\mu_0 H_{EB}$ vs. $E$, measured at $T = 303$ K from individual Kerr loops. Data are taken after initial magnetoelectric annealing of $Cr_2O_3$ (0001)/Pd0.5nm/(Co0.6nmPd1.0nm)$_3$ in axial fields $E = 0.1$ kV/mm and $\mu_0 H = 77.8$ mT. Kerr loops are measured in $E = 0$ after isothermal $E$-field and simultaneous $H$-field exposure of the sample. For a given $\mu_0 H_{EB}$ vs. $E$ curve the magnetic field is constant. The three $\mu_0 H_{EB}$ vs. $E$ data sets correspond to $\mu_0 H = -115$ mT (circles), $\mu_0 H = -154$ mT (triangles), and $\mu_0 H = -229$ mT (squares), respectively. The solid squares in the insets (a) and (b) show the data points of electric switching fields

and corresponding magnetic fields $\mu_0 H = $ -115, -154 and -229 mT. The lines are single parameter fits of the functional form $H = \text{const} / E_c$.

**Methods summary**

Molecular beam epitaxy is used for the sample growth. Ex-situ structural characterization is done by X-ray diffraction techniques. The magnetic characterization is primarily based on the polar magneto-optical Kerr effect and partially performed with the help of a superconducting quantum interference device. The $Cr_2O_3$(0001) surface of the c-$Al_2O_3$/Cr(110)[8 nm]/$Cr_2O_3$(0001) [103 nm] sample was cleaned by ion-beam sputtering and post-sputtering annealing procedures prior to the photoemission measurements. Spin polarized angle resolved photoemission spectra were acquired at the U5UA undulator spherical grating monochromator (SGM) beamline at the National Synchrotron Light Source (NSLS). The electronic structure calculations of the $Cr_2O_3$ (0001) surface were performed using the projected augmented wave (PAW) method[31] as implemented in the Vienna *ab initio* simulation program.[32,33]

**Supplementary Information**

**1. Supplementary Figures and Legends**

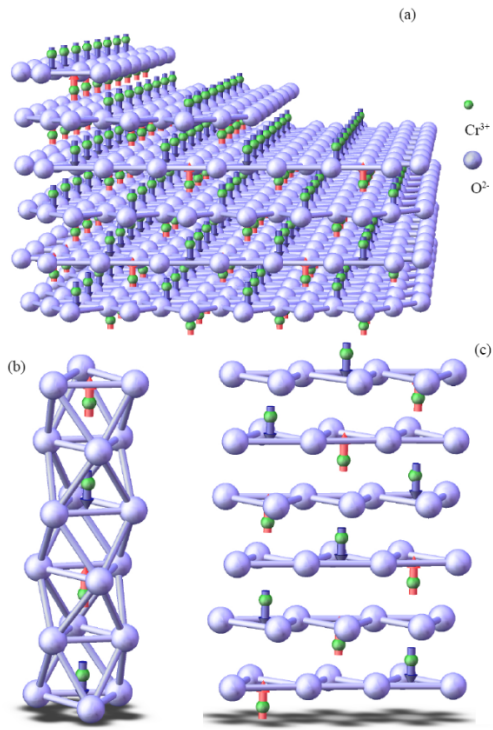

**Supplementary Fig. 1**: (a) Spin structure of a $Cr_2O_3$ single crystal with a stepped (0001) surface is shown for one of its two AF single domain states inverted with respect to Fig.1. Up (red) and down (dark blue) spins of the $Cr^{3+}$ ions (green spheres) point along the *c*-axis. (b) Light blue spheres represent the $O^{2-}$ ions which are arranged in chains of face-sharing distorted octahedra surrounding the $Cr^{3+}$ ions. (c) Shows a more detailed view of the buckled arrangement of the $Cr^{3+}$ ions sandwiched between adjacent oxygen layers.

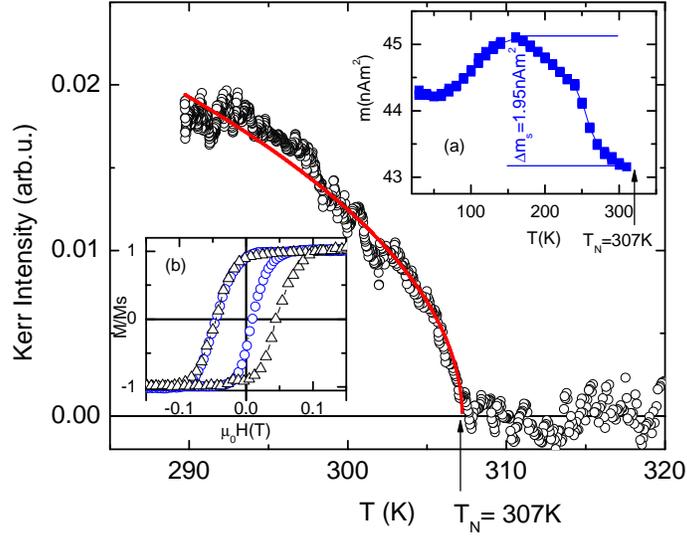

**Supplementary Fig. 2**: Temperature dependence of the polar Kerr signal (circles) of (0001)$Cr_2O_3$(15nm)/Cr(5nm). The Néel temperature $T_N$=307K with the onset of Kerr intensity is indicated by an arrow. The line shows a $\sqrt{T_N - T}$ -fit of the Kerr data. Inset (a) depicts the *T*-dependence of the axial magnetic moment, *m*, of the sample in an applied axial field of $\mu_0 H$ = 0.05T. Horizontal lines indicate the maximum change in *m* vs. *T*. Inset (b) shows the magnetic hysteresis loops of the $Cr_2O_3$(0001)/CoPd exchange bias heterostructure (circles) and the $Al_2O_3$(0001)/CoPd reference sample both measured with the help of polar Kerr magnetometry at *T* = 296 K after field cooling from *T* > $T_N$ in an axial field of $\mu_0 H$ = 0.3 T. The exchange bias loop shift of $\mu_0 H_{EB}$=-16.7 mT is indicated by an arrow.

## 2. Supplementary Methods

The sample investigated in our electrically controlled exchange bias experiments has been grown by molecular beam epitaxy (MBE) at a base pressure of $1.3 \times 10^{-10}$ mbar which increases to $5 \times 10^{-10}$ mbar during growth. A commercial $Cr_2O_3$ single crystal (Crystal GmbH) with an optically flat (0001) surface has been heated to *T* = 723 K and maintained under ultra high vacuum (UHV) condition for 3 hours. Next a Pd[0.5nm]/(Co[0.6nm]Pd[1.0nm])$_3$ multilayer with perpendicular magnetic anisotropy has been MBE grown keeping the $Cr_2O_3$ substrate at *T* = 473 K. Growth rates of 0.87 nm/min and 0.31 nm/min for Pd and Co, respectively have been monitored using a quartz resonator. A reference FM thin film has been deposited by identical means on a diamagnetic sapphire (0001) substrate. A $Cr_2O_3$ (0001) thin film used for spin-resolved photo-emission has been grown by MBE. A Cr (110) seed layer of 8nm thickness has been deposited on the *c*-plane of a sapphire single crystal kept at *T*=573 K during growth.

Subsequently the (0001) $Cr_2O_3$ film has been grown on top of the seed layer by Cr evaporation in the presence of molecular oxygen gas at a partial pressure of $2.6\times10^{-6}$ mbar and substrate temperature of $T$=573 K. The growth procedure described above results in the final structure $c$-$Al_2O_3$/Cr (110)[8nm]/$Cr_2O_3$ (0001)[103nm]. Ex-situ small and wide angle X-ray diffraction has been used for structural analysis of the various samples.

The magnetic characterization was done primarily with the help of polar magneto-optical Kerr effect. Magnetic and electric fields are applied normal to the sample surface. The polarized incident laser beam of wavelength λ=670 nm is reflected from the sample surface in normal incidence geometry. Glan-Thompson polarizers are used for polarizing and analyzing of the light. The reflected beam is periodically modulated between left and right circularly polarized light by the photo-elastic modulator (PEM). Modulation takes place with a frequency of 50 kHz and a phase amplitude of $\varphi_0 = 175^0$ which maximizes the Bessel-function $J_2(\varphi)$. The modulation signal is used as reference signal for a lock-in amplifier. The orthogonal retarder axes of the PEM are perpendicular and parallel aligned to the plane of incidence, respectively. The subsequent analyzer makes an angle of $45^0$ to the retarder axes. The transmitted intensity modulated light is detected by a photodiode providing the input signal to the lock-in amplifier. Its second harmonic Fourier component is proportional to the off-diagonal Fresnel reflection coefficient $r_{sp}$ and, hence, proportional to the magnetization of the sample within the penetration depth of the light beam. In addition a superconducting quantum interference device (Quantum-Design MPMS XL-7) has been used for measurements of the absolute magnetic moment of the $Cr_2O_3$ (0001) sample and its surface in particular.

The $Cr_2O_3$(0001) surface of the $c$-$Al_2O_3$/Cr(110)[8 nm]/$Cr_2O_3$(0001) [103 nm] sample was cleaned by ion-beam sputtering and post-sputtering annealing procedures prior to the photoemission measurements. Spin polarized angle resolved photoemission spectra were acquired at the U5UA undulator spherical grating monochromator (SGM) beamline at the National Synchrotron Light Source (NSLS). Linearly polarized light from an undulator source was monochromatized using a spherical grating monochromator (SGM) and the ultra-high-vacuum photoemission end station was equipped with a commercial angle-resolved hemispherical electron energy analyzer (EA125, Omicron GmbH) and a post electron energy analyzer Mott detector for spin polarization analysis. The spin polarization $P$ for the collected data was determined according to

$$P = \frac{1}{2} \frac{\sqrt{I_L^+ I_R^-} - \sqrt{I_L^+ I_R^-}}{\sqrt{I_L^+ I_R^-} + \sqrt{I_L^+ I_R^-}},$$

where $I_L$ and $I_R$ represent the number of electrons scattered into the left and right channels of the Mott detector, respectively. The spin photoemission spectra were measured repeatedly at $T$=100 K with at an incident photon energy of $h\nu = 58.2$ eV and combined energy resolution of $\Delta E = 120$ meV. Measurements of energy and spin-resolved photoemission can be perturbed by magnetic stray fields. Hence, a magnetoelectric annealing procedure was performed prior to the spin-polarized photoemission measurements using a small applied magnetic and a simultaneously applied electric field is crucial for establishing the antiferromagnetic single domain state.

The electronic structure calculations of the $Cr_2O_3$ (0001) surface were performed using the projected augmented wave (PAW)[31] method as implemented in the *ab initio* total energy and molecular dynamics program VASP (Vienna *ab initio* simulation program).[32,33] The electron-electron interaction is described using the spherically-symmetric version of the LSDA+$U$ method[34] with the parameters $U$ = 4.0 eV and $J$ = 0.58 eV for the Cr 3$d$ electrons, which provide excellent description of the bulk properties of $Cr_2O_3$.[35] The (0001) surface is modeled by a symmetric slab containing eight O layers, 7 bulk-like buckled Cr layers, and a semi-layer of Cr on each surface of the slab. The terminating Cr atoms are placed in the bulk-like positions, which were found to be energetically favorable. The atomic positions were fully optimized while keeping the in-plane lattice translations fixed at their calculated bulk values; the Hellman-Feynman forces were converged to less than 0.01 eV/Å. This optimization and electronic self-consistency were performed using the plane-wave cutoff of 520 eV and a 4×4×1 Monkhorst-Pack $k$-point mesh with a 0.2 eV Gaussian smearing. The site-resolved densities of states were then calculated using a 8×8×1 $k$-point mesh and the modified tetrahedron method. Further details on the surface structure, energetics, and magnetism will be reported in a forthcoming theoretical publication.

### 3. Supplementary discussion

Supplementary Fig. 2 shows the temperature dependence of the local magneto-optical Kerr signal measured in polar reflection geometry in zero applied field on a 15nm $Cr_2O_3$ (0001) thin film. The latter is covered with a protective Cr electrode of 5nm thickness. The incoming laser beam is probing a spot with a lateral size of about 19.6 mm$^2$. The Kerr signal reveals a residual FM moment with a temperature dependence roughly following the power law behavior of a classical order-parameter (line). The onset of a polar Kerr signal at $T < T_N$= 307K indicates a gradually increasing local imbalance between up and down FM surface domains within the finite area probed by the laser spot. The positive or negative surface magnetization of each domain originates from the surface closure spin structure of the AF bulk domains. The smallness and, hence, high noise level of the signal reflects the fact that most of the spontaneous surface magnetization mutually cancels out in a multi domain state. Magnetoelectric annealing of a $Cr_2O_3$ single crystal allows producing a single domain state and a laterally uniform Kerr signal can be measured across the (0001) surface. This global magneto-optical signal of a magnetoelectrically annealed sample has been observed by Krichevtsov et al. in *J. Phys. Condens. Mat.* **5**, 8233 (1993). They discussed their early experimental findings phenomenologically in terms of non-reciprocal light reflection using general symmetry arguments without reference to the intuitive microscopic explanation of a FM surface state evidenced here. The remarkable similarity of the temperature behavior of the Kerr signal measured from $Cr_2O_3$ (0001)/vacuum and a Cr-covered $Cr_2O_3$ (0001) surface implies that the surface magnetic state is robust not only against roughness but maintains also as interface magnetization in bilayer systems. This detail is of significance for our $Cr_2O_3$ (0001)-based exchange bias system.

The FM surface state is further evidenced by quantitative measurements of the axial magnetization of a 200nm $Cr_2O_3$ (0001) thin film covered with a 5nm Pt electrode using superconducting quantum interference magnetometry. The inset (a) of supplementary Fig. 2 displays the temperature dependence of the magnetic moment, *m*, for an axial magnetic field of $\mu_0 H$ = 0.05T applied along the *c*-axis. The change in the magnetic moment between the background value at $T = T_N$ and the maximum value at about $T$ =

260K is $\Delta m = 1.95 \times 10^{-9}$ Am$^2$ (see inset (a) supplementary Fig.2). A straightforward estimate suggests that $\Delta m$ originates from a fully polarized surface layer of Cr$^{3+}$ magnetic moments. With a sample surface of $A = 25 \times 10^{-6}$ m$^2$ and 2 Cr$^{3+}$ ions per surface unit cell one derives an individual Cr$^{3+}$ moment of $2.6 \times 10^{-23}$ Am$^2 = 2.8 \mu_B$ when taking into account an area of $63.9 \times 10^{-20}$ m$^2$ of the hexagonal 1×1 surface mesh. This experimental value of the Cr$^{3+}$ magnetic moment is in excellent agreement with first principles theoretical prediction as well as a basic estimate from Hund's rule. The latter predicts a Cr$^{3+}$ moment of $3\mu_B$ for the $J = S = 3/2$ total angular momentum of the 3d electrons with quenched orbital momentum and Landé-factor $g = 2$. Note that in Cr$_2$O$_3$ (0001) thin films moderate axial magnetic cooling fields are sufficient to favor the FM (0001) surface order without the necessity for magnetoelectric annealing. This indicates that the growth of AF domains stabilizing the surface magnetism starts at the surface. Here symmetry is broken and the spins get easily aligned in an applied field due to the reduced number of AF coupling nearest neighbors. The aligned surface moments act like seeds for the growth of the AF long range order through inhomogeneous nucleation.